\documentclass[final,3p,times,twocolumn]{elsarticle}%{revtex4}%
\usepackage{graphicx}
\usepackage[psamsfonts]{amssymb}
\usepackage{amsmath}

\journal{Nuclear Physics B}

\begin{document}
\begin{frontmatter}

\title{Heavy-Baryon Spectroscopy from Lattice QCD}
\author[UW]{Huey-Wen Lin}
\author[BU]{Saul D. Cohen}
\author[WM,JLAB]{Liuming Liu}
\author[Tata]{Nilmani Mathur}
\author[WM,JLAB]{Kostas Orginos}
\author[WM]{Andre Walker-Loud}

\address[UW]{Department of Physics, University of Washington, Seattle, WA 98195}
\address[BU]{Center for Computational Science, Boston University, Boston, MA 02215}
\address[WM]{Department of Physics, College of William \& Mary, Williamsburg, VA 23187}
\address[JLAB]{Thomas Jefferson National Accelerator Facility, Newport News, VA 23606}
\address[Tata]{Department of Theoretical Physics, Tata Institute of Fundamental Research, Mumbai 400005}

\begin{abstract}
We use a four-dimensional lattice calculation of the full-QCD (quantum chromodynamics, the non-abliean gauge theory of the strong interactions of quarks and gluons) path integrals needed to determine the masses of the charmed and bottom baryons.
In the charm sector, our results are in good agreement with experiment within our systematics, except for the spin-1/2 $\Xi_{cc}$, for which we found the isospin-averaged mass to be $\Xi_{cc}$ to be $3665\pm17\pm14\,{}^{+\,0}_{-\,78}$~MeV.
We predict the mass of the (isospin-averaged) spin-1/2 $\Omega_{cc}$ 
to be $3763\pm19\pm26\,{}^{+\,13}_{-\,79}$~{MeV}.
In the bottom sector,
our results are also in agreement with experimental observations and other lattice calculations within our statistical and systematic errors. In particular, we find the mass of the $\Omega_b$ to be consistent with the recent CDF measurement. We also predict the mass for the as yet unobserved $\Xi^\prime_b$ to be 5955(27)~MeV.
\end{abstract}

\end{frontmatter}

%%%%%%%%%%%%%%%%%%%%%%%%%%%%%%%%%%%%%%%%%%%%%%%%%%%%%%%%%%%%%%%%%%%%%%%%%%%%%%%%%%%%%%%%%%%

Experimental and theoretical studies of charmed and bottom hadrons have been the focus of vigorous research over the last several years. %~\cite{Barberio:2008fa,Voloshin:2007dx,:2007rw,:2007ub}.
In particular, singly and doubly heavy baryon spectroscopy has received significant attention, mainly due to the recent experimental discoveries of both new charmed (SELEX) %~\cite{Mattson:2002vu,Ocherashvili:2004hi}
and bottom baryons by D0 %~\cite{:2007ub}
and CDF. %~\cite{:2007un}.
In addition to these discoveries, there are still many states of heavy and doubly heavy baryons remaining to be discovered. The new Beijing Spectrometer (BES-III), a detector at the recently upgraded Beijing Electron Positron Collider (BEPCII), has great potential for accumulating large numbers of events to help us understand more about charmed hadrons. The antiProton ANnihilation at DArmstadt (PANDA) experiment, a GSI future project, and the LHCb are also expected to provide new results to help experimentally map out the heavy-baryon sector. For these reasons, lattice quantum chromodynamics (QCD) calculations of the spectrum of heavy baryons are now very timely and will play a significant role in providing theoretical first-principles input to the experimental program.

Lattice QCD is a discretized version of four-dimensional continuum QCD. It specializes in studying the strong-coupling regime of QCD, where perturbative approaches converge poorly. As in continuum QCD, we calculate an observable of interest through a path integral:
\begin{multline}
\langle 0| O (\overline{\psi},\psi,A)|0 \rangle =\frac{1}{Z}\int [d A][d \overline{\psi}][d \psi] \\
O (\overline{\psi},\psi,A) e^{i\int\!dx^4{\cal L}_{\rm QCD}(\overline{\psi},\psi,A)},
\end{multline}
where ${\cal L}_{\rm QCD}$ is the sum of the pure-gauge and fermion Lagrangian, $O$ is the operator that gives the correct quantum numbers for our observable, and $Z$ is the partition function of the space-time integral of the QCD Lagrangian. It is straightforward to carry out this path integral numerically within a finite space-time volume and under an ultraviolet cutoff (the lattice spacing); in some physical cases, these approximations to the continuum can dominate the systematic error; we must to carry out calculations using multiple volumes and lattice spacings to extrapolate to the continuum world.

%% sea
In this work we use the ``coarse'' lattice gauge configurations (with lattice spacing $a\approx 0.125$~fm) generated by the MILC Collaboration, %\cite{Bernard:2001av}
in which the asqtad-improved Kogut-Susskind action is used for the fermions in the vacuum.
% light
However, for the valence light quarks (up, down and strange) we use five-dimensional Shamir %~\cite{Shamir:1993zy,Furman:1994ky}
domain-wall fermion propagators %~\cite{Kaplan:1992bt}
with a fifth dimension of extent $L_5=16$ and a mass parameter $M_5=1.7$.
We use hypercubic-smeared gauge links %~\cite{Hasenfratz:2001hp,DeGrand:2002vu,DeGrand:2003in,Durr:2004as}
to minimize the residual chiral symmetry breaking.
The pion masses range from 290 to 597~MeV, and the lattice volume is $20^3\times64$.

For the charm quark we use Fermilab action, %~\cite{ElKhadra:1996mp}
which controls discretization errors of $O((a m_Q)^n)$.
The interpolating operators we use for the $J=1/2$ singly and doubly charmed baryons are
\begin{gather}
\begin{aligned}
\mathcal{O}_{\Lambda_{c}}:&\,\epsilon^{ijk}(q_u^{iT} C\gamma_5 q_d^j)Q_c^k,&
\mathcal{O}_{\Sigma_{c}}:&\,\epsilon^{ijk}(q_u^{iT} C\gamma_5 Q_c^j)q_u^k, \\
\mathcal{O}_{\Xi_{c}}:&\,\epsilon^{ijk}(q_u^{iT} C\gamma_5 q_s^j)Q_c^k,&
\mathcal{O}_{\Omega_{c}}:&\,\epsilon^{ijk}(q_s^{iT} C\gamma_5 Q_c^j)q_s^k,\\
\mathcal{O}_{\Xi_{cc}}:&\,\epsilon^{ijk}(Q_c^{iT} C\gamma_5 q_u^j)Q_c^k,&
\mathcal{O}_{\Omega_{cc}}:&\,\epsilon^{ijk}(Q_c^{iT} C\gamma_5 q_s^j)Q_c^k,
\end{aligned}\nonumber \\
\mathcal{O}_{\Xi^\prime_{c}}:\,\frac{1}{\sqrt{2}} \epsilon^{ijk}\left[(q_u^{iT} C\gamma_5 Q_c^j)q_s^k
	+ (q_s^{iT} C\gamma_5 Q_c^j)q_u^k\right],
\end{gather}
where $q_{u,d,s}$ are the up-, down- and strange-quark fields, and $Q_c$ is the charm-quark field.
Given these actions and operators, we construct two-point functions using the path integral
\begin{equation}
	C_h(t,t_0)=\sum_{\vec{x}}\left\langle \mathcal{O}_h(\vec{x},t)\mathcal{O}_h^\dagger(\vec{x},t_0)\right\rangle ,
\end{equation}
where $\mathcal{O}_h$ is an interpolating operator of the hadron $h$ and $\langle ... \rangle$ indicates an average over the lattice gauge ensemble.
The correlation functions are calculated with gauge-invariant Gaussian-smeared sources and point sinks.
The two-point correlators of the hadrons behave at large time like
\begin{equation}
	C_h(t)=Ae^{-E_0t},
\end{equation}
and the ground-state mass can be obtained from a single-exponential fit.

We use static-quark action (with HYP-smeared links) for the bottom quarks. %
In the static limit, the bottom-quark propagator is reduced to a Wilson line. We can construct the static-light hadron correlation functions by contracting the static-quark line with light-quark propagators to create gauge-invariant correlation functions; the bottom-light baryon correlation functions have the form
\begin{multline}
G_{\Gamma}(\vec{x},t;\vec{x}_0,t_0) =
  \left\langle
    q_f^a(\vec{x},t) \Gamma q_{f^\prime}^b(\vec{x},t)
    \epsilon_{cab} \right.\\\left.
    P^{cc^\prime}(\vec{x},t;\vec{x}_0,t_0)
    \bar{q}_{f^\prime}^{a^\prime}(\vec{x}_0,t_0) \Gamma \bar{q}_f^{b^\prime}(\vec{x}_0,t_0)
    \epsilon_{c^\prime a^\prime b^\prime}
  \right\rangle,
\end{multline}
where $f \in \{u,d,s\}$ is the light-quark flavor index and $P^{cc^\prime}$ is the Wilson line connecting the source and the sink which are separated by time $t$. 
Since the gamma matrix insertions in the baryon correlators are symmetric ($C\gamma_\mu$) or anti-symmetric ($C\gamma_5$), the corresponding flavor wavefunctions are symmetric (sextet) or antisymmetric (antitriplet), respectively. This flavor symmetry structure is automatically enforced without explicit symmetrization or antisymmetrization of the flavor wavefunctions. In the static limit the hyperfine splittings are exactly zero, so we cannot compute such spin splittings. For the physical bottom quark the hyperfine splittings are less than one percent. When comparing to experiment, we identify our states with the lowest-spin physical particles. 
In the case of bottom hadrons, we study the mass splitting between two hadrons $C_{h_1}(t)/C_{h_2}(t)$; this eliminates the unknown additive shift of the mass caused by using static (infinitely massive) quarks.

\begin{figure*}
\includegraphics[width=0.45\textwidth]{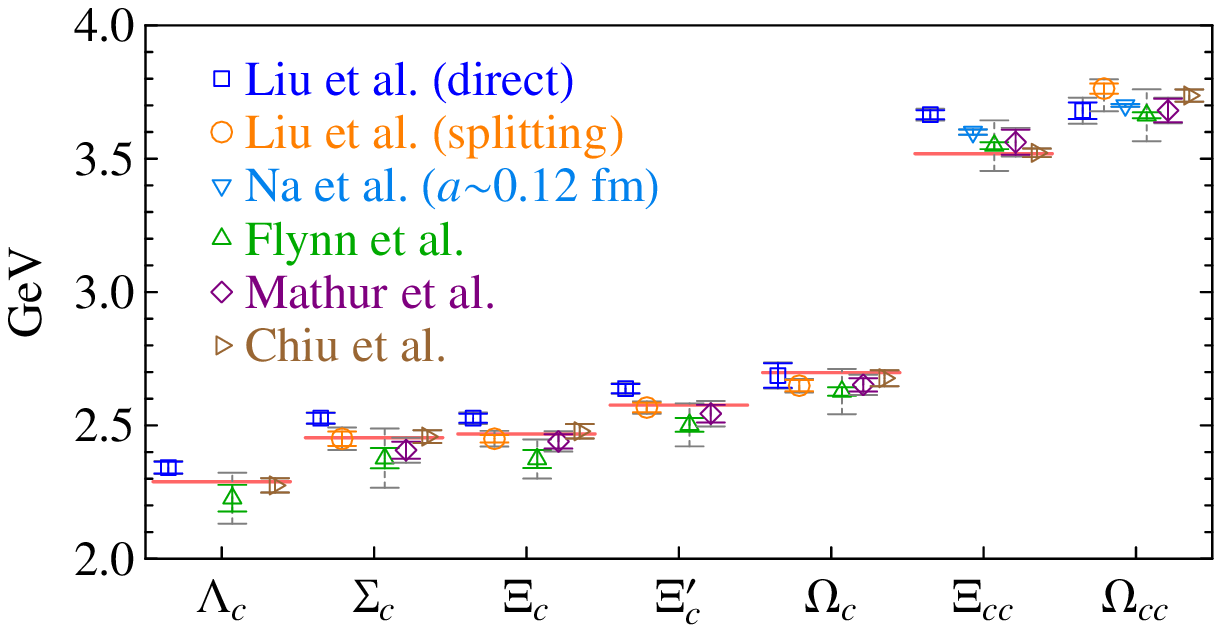}
\includegraphics[width=0.45\textwidth]{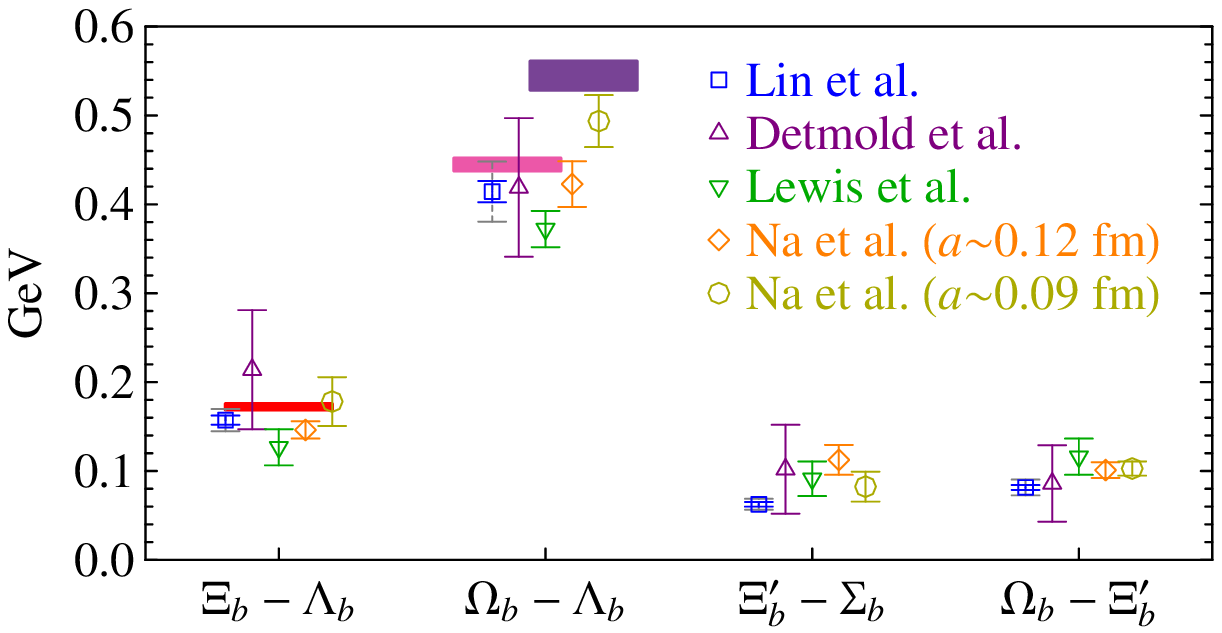}
\caption{\label{fig:all2p1-lat}
(Left) A summary of charmed baryon masses (in GeV) calculated using LQCD.
Our main results~\cite{Liu:2009jc} are the lighter (orange) points taken from splitting extrapolation and the darker (blue) points taken from direct mass extrapolation.
(Right) Comparison of mass splittings with all available 2+1-flavor lattice calculations of bottom baryons. The square (blue) points are extrapolated using the $\Omega$-mass reference scale~\cite{Lin:2009rx}; the solid error bars indicate the statistical error, and the dashed bars indicate the total errors (including the estimated systematics). The solid (red) bars indicate the experimental values given in the PDG, where available. For the $\Omega_b$, we show both the D0 result~\cite{Abazov:2008qm} (upper-right, purple) and the CDF result~\cite{Aaltonen:2009ny} (lower-left, magenta).
}
\end{figure*}

Our results after extrapolation to the physical pion mass are summarized in Fig.~\ref{fig:all2p1-lat}.
We also compare our results with the dynamical lattice charmed-baryon calculation~\cite{Na:2008hz} and quenched calculations~\cite{Chiu:2005zc,Mathur:2002ce,Flynn:2003vz}, shown in Fig.~\ref{fig:all2p1-lat}. Ref.~\cite{Na:2008hz} performed a calculation with staggered light quarks and an interpretation of Fermilab action defining the charm mass by the kinetic mass instead of the rest mass. They observe a significant lattice-spacing dependence in their calculation, which is very different from previous quenched calculations~\cite{Mathur:2002ce}. % with quenched approximation.
Their calculation covers ensembles at more than one lattice spacing, but their results are not as precisely determined statistically. Their uncertainties given here are merely the statistical ones; no systematic uncertainties are given yet.

We compare the doubly charmed baryons with predictions from theoretical models;
although the SELEX Collaboration has reported a first observation of doubly charmed baryons, searches by the BaBar, %~\cite{Aubert:2006qw},
Belle %~\cite{Chistov:2006zj}
and Focus %~\cite{Ratti:2003ez}
Collaborations have not confirmed their results. This makes it interesting to look back to theory to see where the various predictions lie.
We compute the mass of $\Xi_{cc}$ to be $3665\pm17\pm14\,{}^{+\,0}_{-\,78}$~MeV, which is higher than what SELEX observed.

We compare with a selection of other theoretical results, such as a recent quark-model calculation, %~\cite{Roberts:2007ni}
relativistic three-quark model, %~\cite{Martynenko:2007je}
the relativistic quark model, %~\cite{Ebert:2002ig}
the heavy quark effective theory %~\cite{Korner:1994nh}
and the Feynman-Hellmann theorem.  %~\cite{Roncaglia:1995az}.
Most theoretical results suggest that the $\Xi_{cc}$ is about 100--200~MeV higher.
The $\Omega_{cc}$ mass prediction made by this work is $3763\pm19\pm26\,{}^{+\,13}_{-\,79}$~MeV, and the overall theoretical expectation is for the $\Omega_{cc}$ to be 3650--3850~MeV. We hope that upcoming experiments will be able to resolve these mysteries of doubly charmed baryons. 
In the future we plan to extend these calculations to a second lattice spacing, address systematics due to continuum extrapolation, and extend our spectroscopy to spin-3/2
states.

Our results for bottom baryons are shown on right-hand-side of Fig.~\ref{fig:all2p1-lat}, along with other 2+1-flavor works\cite{Na:2008hz,Lewis:2008fu,Detmold:2008ww}.
All four calculations include the splittings $\Xi_b-\Lambda_b$ and $\Sigma_b-\Lambda_b$; we calculate the additional splittings $\Xi_b^\prime-\Sigma_b$ and $\Omega_b-\Xi_b^\prime$ in order to make direct comparisons. 
We again see good agreement amongst all lattice calculations, and mild scaling in the Na~et~al. results. Agreement with experimental values as given by the PDG is fairly good. The discrepancy in the $\Sigma_b-\Lambda_b$ splitting may be due to discretization effects, as suggested by the trend visible in the Na~et~al. results.

Our results agree well with experiment in the cases of known mass splittings. In the case of $\Omega_b$ our calculation is in agreement with the recent CDF result and several standard deviations away from the D0 result. Such a conclusion holds for all other lattice calculations of the $\Omega_b$ mass. Our results for the $\Xi_b^\prime$ mass splittings are a prediction since $\Xi_b^\prime$ has not yet been observed. Using the splitting from the $\Lambda_b$ and summing with the experimental value of that baryon, we predict the mass of the $\Xi_b^\prime$ to be 5955(27)~MeV, where the error given combines in quadrature all our errors. 
For more details about the lattice calculation (including the chiral extrapolations to physical pion mass and systematic error estimations) and references, we refer readers to Refs.~\cite{Liu:2009jc,Lin:2009rx}.

In the future we plan to extend these calculations using the anisotropic gauge ensembles generated by the Hadron Spectrum Collaboration (HSC)\cite{Lin:2008pr,Edwards:2008ja}.
The anisotropy (giving finer temporal lattice spacing while keeping spatial one large enough to keep the finite-volume systematic error small) will provide finer resolution not only to improve the signal quality but also to reliably probe the excited states.
In the charm sector, we will extend the calculations to $J=3/2$ baryons and adopt a relativistic improved action for the bottom quark to get rid of the mass redundancy in the bottom sector. 
A recent paper by HSC~\cite{Peardon:2009gh} demonstrates a new technique, optimized for complicated operators, which may further improve the calculations and in particular allow for extraction of the excited-state spectrum of both the charm- and bottom-baryon sectors.

{\small{\it Acknowledgments} HL is supported by the U.S. Dept. of Energy under Grant No. DE-FG03-97ER4014. HL thanks
European Physical Society (EPS) for awarding the conference grants to attend the 2009 Conference on Computational Physics. These calculations were performed using the Chroma software suite\cite{Edwards:2004sx}.

%\bibliographystyle{elsarticle-num}
%\bibliography{bottom,charmed}
\end{document}